\documentclass[12pt]{article}

\usepackage[english]{babel}

\usepackage[a4paper,top=3cm,bottom=3cm,left=3cm,right=3cm,marginparwidth=1.75cm]{geometry}

\usepackage{graphicx}
\usepackage[colorlinks=true, allcolors=blue]{hyperref}

\usepackage{amsmath,amsfonts,amssymb,amsthm,mathtools}

\title{Neural Network Tuning of FSMPC for Drives}
\author{Juana M. Martínez-Heredia, José L. Mora}

\begin{document}
\maketitle

%
\begin{abstract}
This preprint  presents a neural network tuner for the finite state model predictive control of an induction motor. The tuner deals with the parameters of the controllers in the speed loop and in the stator current loop. The results are assessed using a five phase machine in an experimental setup. Data for the neural network training is obtained from the experiments using step tests.
\end{abstract}

%
\section{Introduction}
Direct digital predictive control of the discrete states of the power converter has been recently proposed for many applications. The method has been termed Finite State Model Predictive Control (FSMPC) \cite{arahal2016harmonic}. The benefits of FSMPC lie mainly in the added flexibility that is used to cope with different objectives. Also, an increase in bandwidth in the stator current control has been reported \cite{Levi_TIE2014comparative} due to the elimination of the modulation block \cite{colodro2009analog}.

Previous papers of FSMPC for multi-phase systems have recognized the problem of CF tuning  with regard to stator current control \cite{liu2017overview}. Among the different proposals, one can find adaptive methods and table-based ones. Regarding adaptive methods, the idea of model reference has been used in \cite{arahal2021adaptive,arahal2025model}. Similarly, in \cite{makhamreh2020lyapunov} a Lyapunov-based design is proposed.

Neural Networks (NN) have been extensively used for control systems \cite{arahal1997nonlinear,zhang2025predictor}. Their capacity for interpolation can be exploited to provide the control parameters for each operating regime \cite{colodro1996cellular}. The NN approach is often used with a previous off-line learning phase \cite{hammoud2022learning,wai2001adaptive}. The method has been applied to the speed loop. For instance, in \cite{lu2014multivariable}, the speed PI is replaced by a  fuzzy or neural system. Similar ideas have been used in  \cite{lu2021speed} for PID tuning and in \cite{kanungo2023design}, where a Wavelet-based adaptive method is used.

In the case of  electrical rotating machines, the speed and current loops should be considered. This can be seen in \cite{sangar2024improved}, where an adaptive system replaces the speed controller. This allows for a nonlinear control adjusted to each   operating point. The control of  all motor variables is proposed in \cite{wrobel2020model,colodro2014linearity,devin2023dynamic}. In this case, NN have been used to obtain acurate models to be used in FSMPC \cite{hanke2021comparison}. Another approach is to use the NN to reproduce the behavior of a predictive controller \cite{abu2022deep,karimi2023continuous,colodro2011frequency}.

This preprint  presents a neural network tuner for the finite state model predictive control of an induction motor. The tuner deals with the parameters of the controllers in the speed loop and in the stator current loop. The results are assessed using a five phase machine in an experimental setup.

%
\section{Control scheme}
\begin{figure}
\centering
  \includegraphics[width=13cm]{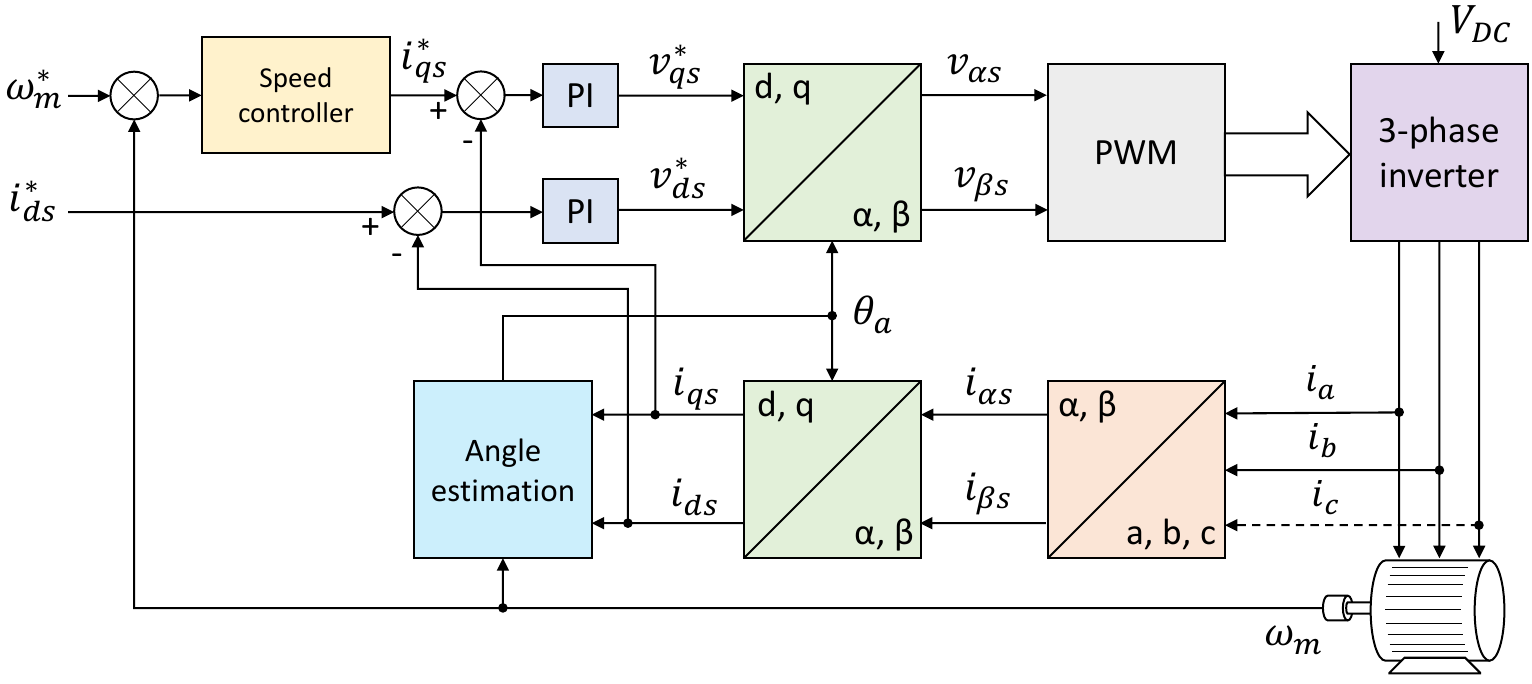}
  \caption{Diagram of the induction motor control.}\label{fig_ifoc_scheme}
\end{figure}

The motor control scheme is based on ideas of IFOC and FSMPC. This is illustrated in Fig. \ref{fig_ifoc_scheme}, where a multi phase induction motor of 5 phases is considered. 

The voltage source inverter (VSI) can produce $2^5$ possible voltages that supply the 5 motor phases. This corresponds to the states of its discrete switches. If $u_i$ denotes the state of the switch of the $i-th$ leg then the stator phase voltages are $v_i$ as $V_{ph} = T U$, where $V_{ph}=( v_1, ..., v_5)$, $U=( u_1, ..., u_5)$ and $T$ is given by

\begin{equation}
T=\frac{V_{DC}}{5}
\begin{pmatrix*}[r]
~4&-1&-1&-1&-1\\
-1&~4&-1&-1&-1\\
-1&-1&~4&-1&-1\\
-1&-1&-1&~4&-1\\
-1&-1&-1&-1&~4
\end{pmatrix*},
\end{equation}

\noindent where $V_{DC}$ is the DC-link voltage. Thse voltages can be projected to the torque producing plane   ($\alpha-\beta$) and to the harmon ic plane  ($x-y$) using the transformation matrix

\begin{equation}\label{eq_matrizM}
M=\frac{2}{5} 
\begin{pmatrix*}[r]
1&  \gamma^c_1&  \gamma^c_2&  \gamma^c_3&  \gamma^c_4\\
0&  \gamma^s_1&  \gamma^s_2&  \gamma^s_3&  \gamma^s_4\\
1&  \gamma^c_2&  \gamma^c_4&  c_\vartheta& \gamma^c_3\\
0&  \gamma^s_2&  \gamma^s_4&  \gamma^s_1&  \gamma^s_3\\
1/2&1/2&1/2&1/2&1/2
\end{pmatrix*},
\end{equation}

\noindent where $\gamma^c_h = \cos{h \vartheta}$, $\gamma^s_h = \sin{h \vartheta}$,  and $\vartheta = {2\pi}/5$ (rad). Quantities in $\alpha-\beta-x-y$ are obtained as $V_{\alpha\beta xy} = M V_{ph}$. For the realization of the Indirect Field Oriented Control (IFOC) scheme, quantities in the rotating $d-q$ reference frame are also needed. The $d$ component deals with flux production and the $q$ component with torque. The Park rotation matrix is then used to convert the $\alpha-\beta$ axes into $d-q$, so $V_{d-q} = D V_{\alpha\beta xy}$, with 

\begin{equation}\label{eq_park}
  D = 
\begin{pmatrix*}[r]
   ~\cos ~ \sigma  & \sin ~ \sigma \\
   -\sin ~ \sigma &  \cos ~ \sigma\\
  \end{pmatrix*},
\end{equation}

\noindent where the flux position $\sigma$ is obtained as $\sigma = \int \omega_e dt$.

The $d-q$ stator currents  reference are $i_{dq}^* = ( i_{d}^*, i_{q}^*)$. The $d$ component is fixed to produce a constant flux, the $q$ component produces torque to drive the mechanical speed control error to zero and is computed as 

\begin{align}\label{eq_pi_veloc}
i^{*}_{q}(t) = k_p \cdot e_\omega(t) + k_i \int_0^t e_\omega( \tau) d \tau
\end{align}
\noindent where $e_\omega(t) = \omega^*(t) - \omega^e(t)$ is the velocity error.

The value $i_{dq}^*$ must be  transformed to $\alpha-\beta-x-y$ for stator current control, resulting in 
 $i^*_{\alpha}(t)=I^* \sin{ \omega_e t}$, $i^*_{\beta}(t)=I^* \cos{ \omega_e t}$, $i^*_{x}(t)=0$, $i^*_{y}(t)=0$ with $I^* = \| i_{dq}^* \|$.

The FSMPC part is responsible for stator current tracking of $i^*_{\alpha\beta xy}$. For this, a model-based scheme is used where the control action is determined minimizing a cost function. The model provides the one-step ahead prediction as

\begin{equation}\label{eq_p1p}
\hat{i}_{\alpha\beta xy}(k+1) = \Phi ( \omega) i_{\alpha\beta xy}(k) +  \Psi U (k).
\end{equation}

In order to cope with the delay in computations a second prediction is used. This is obtained as $\hat{i}_{\alpha\beta xy}(k+2) = \Phi ( \omega) \hat{i}_{\alpha\beta xy}(k+1) +  \Psi U(k+1)$.

The control action  $U(k+1)$ is computed at discrete time $k$ minimizing a cost function $J$ consisting of several terms. Penalties are placed  of predicted errors in $\alpha-\beta$, $x-y$ and the number of switch changes are typically used, resulting in  

\begin{equation}\label{eq_fcoste}
  J = \| \hat{E}_{\alpha\beta} (k+2) \|^2 +  \lambda_{xy} \| \hat{E}_{xy} (k+2) \|^2 + \lambda_{sc} SC(k+1)
\end{equation}

\noindent where $\| \hat{E}_{\alpha\beta} \|^2$ is the quadratic deviation of predictions from reference in $\alpha-\beta$ plane, $\| \hat{E}_{xy} \|^2$ is the quadratic deviation of predictions from reference in $x-y$ plane and $SC$ is the number of switch changes in the VSI:

\begin{equation}\label{eq_def_NC}
  SC(k+1) = \sum_{h=1}^{5} | u_h(k+1) - u_h(k) |
\end{equation}

The weighting factors of the cost function are parameters $\lambda_{xy}$ and $\lambda_{sc}$. These parameters have an influence on the selected control action on each sampling period, as a result they influence the performance indices obtained as will be shown in Section 2.1.

\subsection{Performance indicators}
The performance indicators considered include the mechanical, electro-magnetic and electronic variables listed and defined below. Each indicator is given a number $i$, so that it can be referred to as $\pi_i$.

\begin{itemize}
    \item Speed overshoot ($\pi_1=PO$), defined in (\ref{eq_def_PO}).
    \item Speed rise time($\pi_2=T_r$), defined in (\ref{eq_def_Tr}).
    \item Integral Time Absolute Error ($\pi_3=ITAE$), defined in (\ref{eq_def_ITAE}).
    \item Torque ripple ($\pi_4=R_t$), defined in (\ref{eq_def_torqueripple}).
    \item Harmonic content ($\pi_5=E_{xy}$), defined in (\ref{eq_def_exy}). 
    \item Average switching frequency ($\pi_6=ASF$), defined in (\ref{eq_def_ASF}).
\end{itemize}

The mathematical expressions of the performance indicators are:
\begin{eqnarray} \label{eq_def_PO}
  PO &=& 100 \cdot \frac{ \max{ \omega} - \omega^* }{\omega^*}  \\ \label{eq_def_Tr}
  T_r &=& \underset{t \ge 0}{\operatorname{argmin}}  ~  \omega^*(t) - \omega(t) \\ \label{eq_def_ITAE}
  ITAE &=&  \frac{1}{N}  \sum_{k=1}^{N} 
  \frac{ \| \omega^*(k)  - \omega(k)  \|}{ \omega^*(k) } k \\ \label{eq_def_torqueripple}
  R_t &=& \sqrt{ \frac{1}{N}  \sum_{k=1}^{N} \left(   T^*(k)  - T(k)  \right)^2 }  \\  \label{eq_def_exy}
  E_{xy} &=& \sqrt{ \frac{1}{N} \sum_{k=1}^{N} e_{xy}^2 (k) } \\ \label{eq_def_ASF}
 ASF &=& \frac{1}{5 N  T_s } \sum_{k=1}^{N} SC(k)
\end{eqnarray}

\subsection{Laboratory setup}
\begin{figure}
\centering
  \includegraphics[width=10cm]{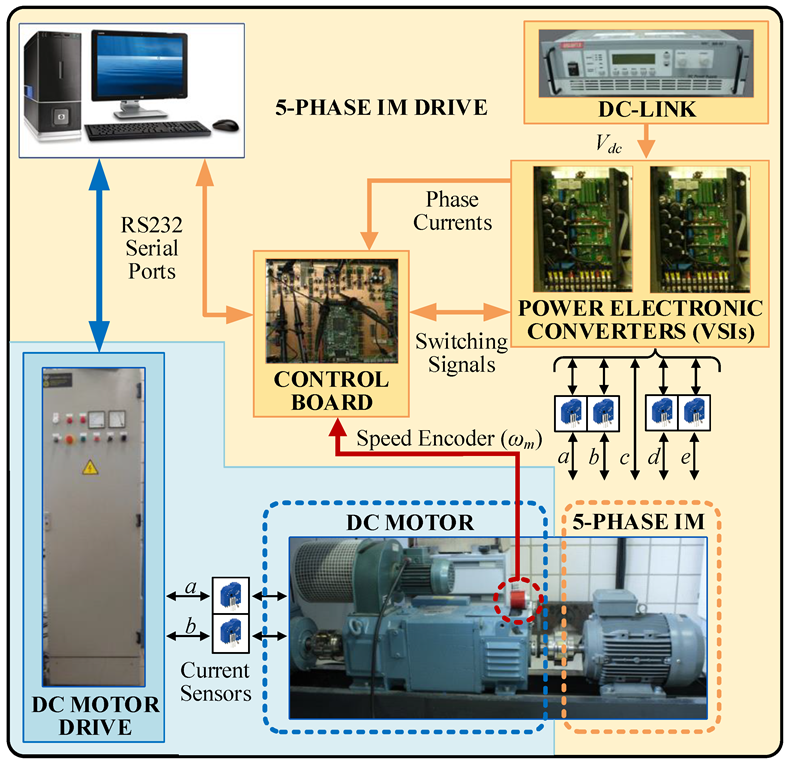}
  \caption{Elements of the experimental setup.}\label{fig_bancada}
\end{figure}

The experimental setup  includes  (see Fig. \ref{fig_bancada})  a 5pIM, with  parameters shown in Table \ref{tab_Parameters}, a power converter using two SEMIKRON SKS 22F modules powered by a 300V DC power supply, a MSK28335 board including a TMS320F28335 Digital Signal Processor (DSP)  where the control program is run in real-time. 

The IFOC scheme of Fig. \ref{fig_ifoc_scheme} is used for the double-loop control. The mechanical speed is sensed using a GHM510296R/2500 encoder. An interruption routine of the DSP is used to compute the mechanical speed from incoming encoder pulses. Regarding the inner control loop, Hall effect sensors (LH25-NP) are used to measure the stator phase currents. 

The setup has the ability of testing the controller with some independence of the mechanical load characteristic curve thanks to a co-axial  DC motor that is used to generate an independent opposing torque load ($T_L$) for the tests. 

\begin{table}[!t]
	\renewcommand{\arraystretch}{1.3}
	\caption{Parameters of the Experimental 5pIM}
	\centering
	\label{tab_Parameters}
		\begin{tabular}{lcc}
			\hline\hline \\[-3mm]
 Parameter & Value & Unit \\[1.6ex] \hline
 Stator resistance, $R_s$           &   12.85 &   $\Omega$  \\
 Rotor resistance, $R_r$            &	4.80 &   $\Omega$  \\ 
 Stator leakage inductance, $L_{ls}$&   79.93 &   mH \\
 Rotor leakage inductance, $L_{lr}$ &   79.93 &   mH \\
 Mutual inductance, $L_M$           &  681.7  &   mH \\
 Rotational inertia, $J_m$          &    0.02 &   kg m$^2$ \\
 Number of pairs of poles, $P$      &  3      & - \\
			\hline\hline
		\end{tabular}
\end{table}

\section{NN tuning}
The proposed method  consists of an Artificial Neural Network (ANN) that provides the optimal PI controller parameters and optimal FSMPC weighting factors according to the actual speed $\omega$ and its reference $\omega^*$. Denoting the set of parameter as $\theta$

\begin{equation}\label{eq_def_theta}
  \theta = \left(  k_p,  k_i,  \lambda_{xy}, \lambda_{sc}   \right),
\end{equation}

\noindent the ANN task is to compute $\theta^* = \mathrm{f}( \omega, \omega^*)$.

The ANN is chosen because of its universal approximation property. Other approximators could be used as well. The simplicity of the perceptron facilitates its implementation in many control hardware platforms such as Field Programmable Gate Array (FPGA). The network architecture is a multilayer perceptron with sigmoidal activation in the hidden units, linear activation in the output units and bias connections to all units. The network size is determined by cross-validation. In addition, network training uses early stopping and a final phase of validation. Training data is gathered by selecting the most appropriate set of parameters for both, the speed-loop PI and the current-loop FSMPC in the way that is presented below.


In the case considered here,  overshoot and VSI switching frequency are given priority. This case can be tackled with the help of the following optimization problem 

\begin{equation}
\begin{aligned} 
    &\underset{ \theta}{\operatorname{min}} && c_2 \pi_2 + c_3 \pi_3 + c_4 \pi_4 + c_5 \pi_5  \\
    &\textrm{s.t.} && \pi_1 \leq U_{PO} \\
    &              && \pi_6 \leq U_{ASF}.
\label{eq_optimz_problem}
\end{aligned}
\end{equation}

The function being optimized $\Xi = c_2 \pi_2 + c_3 \pi_3 + c_4 \pi_4 + c_5 \pi_5 $ is a combination of objectives.  Coefficients $c_i$ must be designed according to the relative importance of each performance index. Also, those performance indices considered most important are treated as constraints. This is the case, in (\ref{eq_optimz_problem}), of $PO <  U_{PO}$ and $ASF < U_{ASF}$. The first constraint ensures that the overshoot will be below limits. The second constraint ensures that the VSI will not overheat and damage due to fast commutations. Please notice that other optimization problems are possible interchanging constraints and terms in $\Xi$. In this manner it is possible to ensure some operational values (constraints) while minimizing the rest of performance indicators (index $\Xi$).

\subsection{Data gathering and training}
Supervised learning is used to obtain an ANN model that provides $\theta$ as a function of $x=(\omega, \omega^*)$. A data-set of input/output values is obtained by minimizing $\Xi$ for a collection of operating points. The minimization is done considering experimental data where pairs of $\Pi$ values are gathered for each tentative $\theta$ and operating point.

The experimental $\Pi$ values have been gathered by performing step tests on the experimental setup. This task should confront the fact that a large number of $\theta$ combinations is possible. Think for instance in  a grid dividing each $\theta_i$ into ten parts, then $10^4$ combinations are possible for vector $\theta$. This number is affordable for simulation but it is prohibitive for an experimental platform. Some techniques have been devised to reduce the number of experimental trials to obtain the training data.

\begin{enumerate}
    \item The outer PI loop can be roughly tuned using a simplified dynamical model corresponding to a second order transfer function. This provides a set of initial guesses for $\theta_1$ and $\theta_2$ that are later refined. 
    \item Similarly, an initial guess for the WF parameters of the CF can be obtained from previous works on the IM as shown in \cite{arahal2021adaptive}.  This provides a set of initial guesses for $\theta_3$ and $\theta_4$.
    \item Finally, instead of the extensive exploration of $\theta$ space featured in previous works, a gradient descent is used to produce new combinations from existing ones. In this way the experimental setup can drive itself in the data gathering task. This is similar to the strategy for real-time WF selection proposed in \cite{arahal2021adaptive} for a six-phase IM.
\end{enumerate}

\begin{figure}
\centering
  \includegraphics[width=11cm]{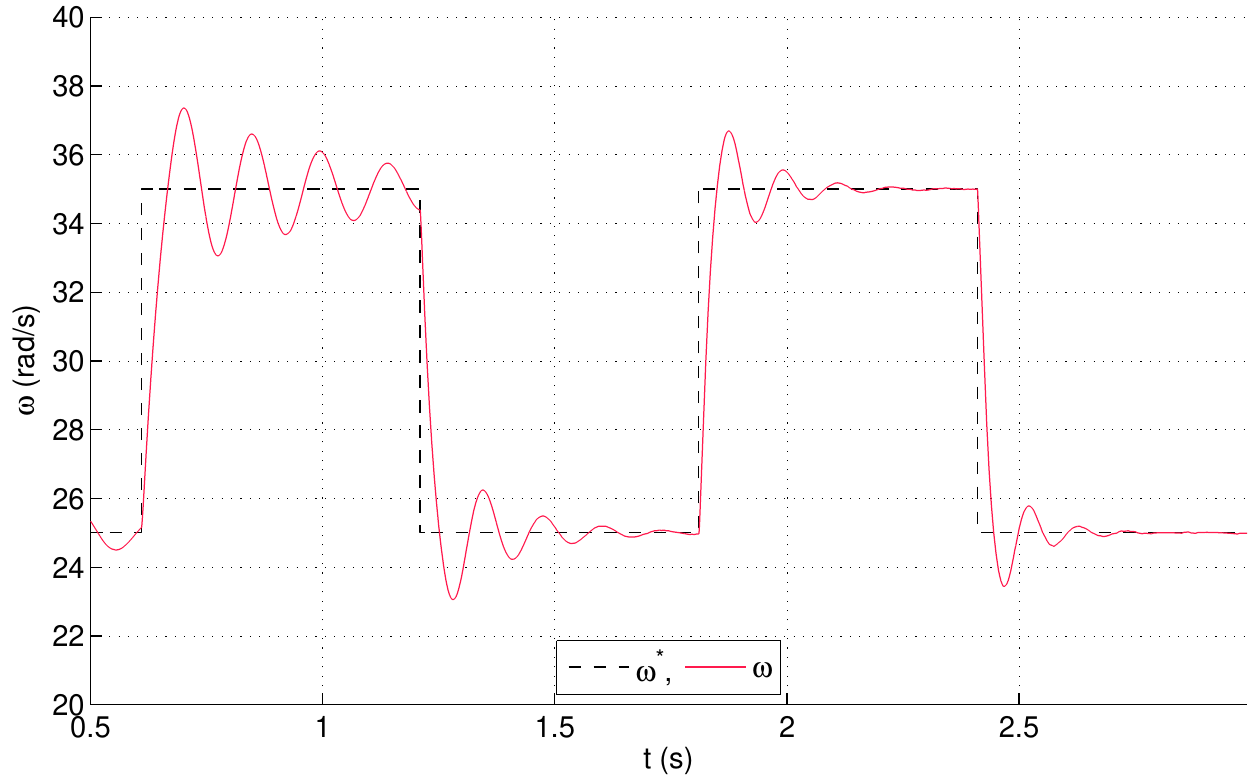}  
  \caption{Example of a test for data gathering.}\label{fig_tray_training}
\end{figure}

The step tests have the form depicted in Fig. \ref{fig_tray_training} where a set of step changes are introduced in $\omega^*$ for a particular operating point. The controller tuning is initially set to $\theta^0$ and later changed after each step. This produces a convergence of the performance indicators towards appropriate values. The procedure is repeated for all operating points in the IM's range. Each test requires less than a minute to be completed, so in a few minutes the whole operational range is covered.

With this strategy a set of input/output pairs is gathered for ANN training.

The network size is selected using the usual validation set and early stopping. The actual training is done using the Matlab Toolbox for shallow networks.

%
\section{Conclusion}
This preprint presents a neural network solution for the on-line tuning of a motor controller. The tuner provides the parameters of the controllers in the speed loop and in the stator current loop. The results are assessed using a five phase machine in an experimental setup. Data for the neural network training is obtained from the experiments using step tests.

%
\bibliographystyle{unsrt}
\bibliography{sample}

\end{document}